\documentclass[aps, prl, reprint]{revtex4-1}

\usepackage{graphicx, epsfig, array, subfigure, epstopdf}
\usepackage[colorlinks=true, citecolor=blue, linkcolor=blue, urlcolor=blue]{hyperref}
\usepackage[colorinlistoftodos]{todonotes}

\begin{document}

\title{Understanding Quality Factor Degradation in Superconducting Niobium Cavities at Low Microwave Field Amplitudes}
\thanks{This work was supported by the US Department of Energy, Offices of High Energy and Nuclear Physics.}%
\author{A. Romanenko} 
\email{aroman@fnal.gov}
\affiliation{Fermi National Accelerator Laboratory, Batavia, IL 60510, USA }
\author{D. I. Schuster}
\email{david.schuster@uchicago.edu}
\affiliation{The James Franck Institute and Department of Physics, University of Chicago, Chicago, Illinois 60637, USA}

\date{\today}

\begin{abstract}
In niobium superconducting radio frequency (SRF) cavities for particle acceleration a decrease of the quality factor at lower fields - a so called \emph{low field Q slope or LFQS} - has been a long-standing unexplained effect. By extending the high $Q$ measurement techniques to ultralow fields we discover two previously unknown features of the effect: i) saturation at rf fields lower than $E_\mathrm{acc} \sim 0.1$~MV/m; ii) strong degradation enhancement by growing thicker niobium pentoxide. Our findings suggest that the LFQS may be caused by the two level systems in the natural niobium oxide on the inner cavity surface, thereby identifying a new source of residual resistance and providing guidance for potential non-accelerator low field applications of SRF cavities.
\end{abstract}


\maketitle

Modern and planned state-of-the-art particle accelerators employ hundreds or thousands of three-dimensional superconducting radio frequency (SRF) niobium cavities~\cite{Padamsee_Ann_Rev_Nucl_2014, Padamsee_Review_SUST_2001} for particle acceleration. In operation, a beam of charged particles (e.g. electrons, positrons, protons, heavy ions) is accelerated by the electric field along the axis of the cavity.  The phase of the field is such that particles always see an accelerating field along their trajectories. Maintaining the large electromagnetic fields inside cavities leads to dissipation, and - compared to normal conducting technology - SRF cavities provide an extremely low power consumption thereby permitting continuous wave (CW) operation as well as enabling superior beam quality.

Physics and technology of SRF cavities has progressed rapidly over the years~\cite{Hasan_book2}, currently allowing unprecedented intrinsic quality factors $Q>2\times10^{11}$ to be attained up to very high rf fields of $E_\mathrm{acc}>$~20~MV/m~\cite{Romanenko_APL_2014}. These advances were achieved by novel surface preparation techniques such as nitrogen doping~\cite{Grassellino_SUST_2013}, and special cooldown procedures to eliminate the residual resistance contribution from trapped DC magnetic flux~\cite{Romanenko_JAP_2014}. These recent findings have translated into significant increases (factor of $>$2-3) in the efficiency of CW particle accelerators (e.g. LCLS-II at SLAC) operated at medium rf accelerating fields up to about 20~MV/m. 

One of the remaining unexplained phenomena in gigahertz range SRF cavities is a strong decrease of quality factor ($Q$) at low rf fields $E_\mathrm{acc} \lesssim 5$~MV/m - the so called ``low field Q-slope'' (LFQS). Reported experimental investigations~\cite{Ciovati_JAP_2004, Visentin_ICFA_2006, Padamsee_Ann_Rev_Nucl_2014} showed a continuous decrease of $Q$ down to $\sim 0.2$~MV/m, the lowest field explored. Most recent studies~\cite{Romanenko_Rs_B_APL_2013} indicate that the increase in average surface resistance (decrease in $Q$) in LFQS does not come from the thermally excited quasiparticle contribution described by Mattis and Bardeen~\cite{Mattis_Surf_Res_1958}, but is a part of the residual surface resistance contribution. The residual resistance currently sets the limit to the maximum possible SRF cavity quality factors~\cite{Gurevich_SUST_2017}, and plays the dominant role for sub-gigahertz range SRF-based accelerators. Understanding the physics of all the mechanisms behind residual resistance is among the major remaining challenges for further SRF progress. 

In addition to the physics of residual resistance, understanding of the LFQS has recently acquired strong practical cross-discipline interest as a range of potential non-accelerating applications of high $Q$ SRF cavities emerged in particle physics~\cite{Jaeckel_PhysLettB_2008}, quantum computing~\cite{Paik_PRL_2011, Paik_PRL_2016, Reagor_PRB_2016}, astrophysics~\cite{Zmuidzinas_ARCMP_2012}, superconducting parametric conversion~\cite{Reece_NIMA_1986}, and gravitational wave detection~\cite{Caves_PhysLettB_1979, Pegoraro_JPhysA_1978}, for which operation in the limit of very low rf fields (down to single photon) and/or temperatures ($T \lesssim$~25~mK) is of interest. The primary interest is due to the high potential of SRF cavities with $Q > 10^{11}$, as compared to maximum reported quality factors of other 3D-resonators in this regime of $Q \sim 10^{8}$~\cite{Reagor_PRB_2016}. The obvious need is then to understand how far down will the $Q$ of SRF cavities drop at ultra low fields due to the LFQS, which requires direct experimental probing. Understanding of the physics of the LFQS will then be of crucial importance for any further surface optimization.

There have been two models of the LFQS discussed in the literature: the first model~\cite{Halbritter_SRF_80} postulated the existence of niobium suboxide clusters within the penetration depth, while the second one~\cite{Palmieri_QSlopes_SRF2005} suggested that the niobium penetration depth can be treated as a two-layer superconductor with the topmost superconductor having the rf field-dependent penetration depth.

In this Letter we report the first $Q$ measurements in the extended accelerating rf field range down to $\sim10^{-5}$~MV/m, which indicate that LFQS may be a form of dielectric loss, rather than conductance loss as hypothesized previously.  We studied a large set of bulk niobium 1.3~GHz SRF cavities of elliptical shape and different surface treatments, which reveal the saturation in the decrease of the $Q$ factor (low field $Q$ slope) below $E_\mathrm{acc} \sim 0.1$~MV/m. Growing a thicker oxide on the rf surface of the cavity leads to a strongly enhanced low field dissipation, identifying oxide as a primary contributor to the effect. Combined, these two findings suggest that the low field $Q$ slope in bulk niobium SRF cavities, which eluded solid explanation for more than two decades, may be similar in nature to that found in planar resonators~\cite{Martinis_PRL_2005, Gao_APL_2008, Kaiser_SUST_2010, Muller_arXiv_2017}, i.e. caused by the two-level systems (TLS) present in the native niobium oxide Nb$_2$O$_5$ covering the inner resonator surface. Our experimental data is also not compatible with the previously proposed LFQS models. Furthermore, the residual resistance at higher rf fields is also changed by anodizing, highlighting the oxide contribution at all fields.

\begin{figure}[htb]
 \includegraphics[width=\linewidth]{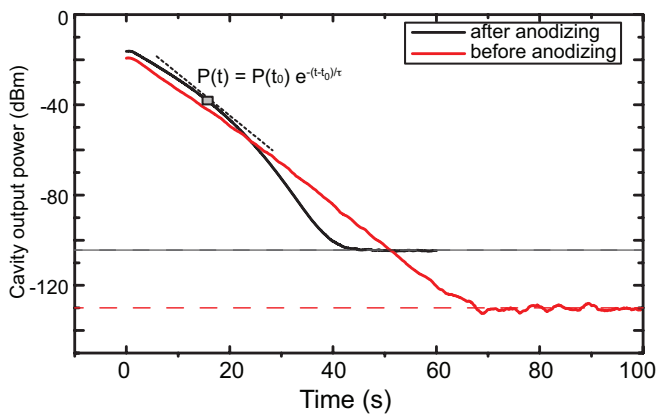}
\caption{\label{fig:Pt}Decay of cavity output power upon turning the RF source off. At each moment in time $t_0$, time decay can be characterized by the ``instantaneous'' time constant $\tau(t_0)$, from which the loaded quality factor $Q_L$ is obtained. Dashed lines reflect the noise floor in the respective configurations.}
\end{figure}

\begin{figure}[htb]
 \includegraphics[width=\linewidth]{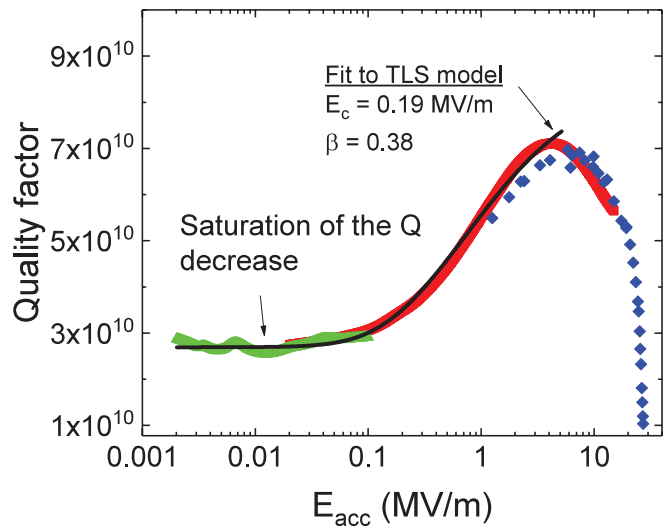}
\caption{\label{fig:Saturation}Q(E) in the full field range for electropolished 1-cell cavity obtained by concatenating overlapping CW and single shot decay measurements. RBW=10~kHz has been used for the red curve to compensate for the Lorentz force detuning at large rf fields. The fit to TLS model (Eq.~\ref{eq:TLS}) is shown as well.}
\end{figure}

\begin{figure}[htb]
 \includegraphics[width=\linewidth]{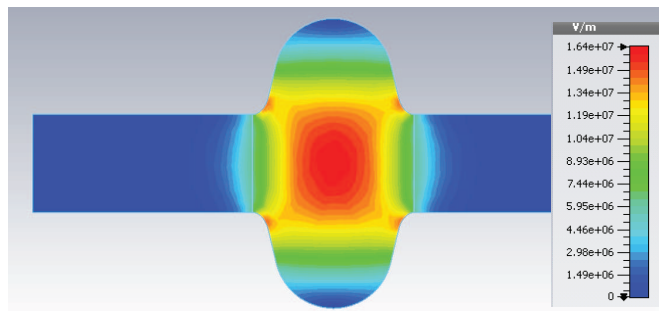}
\caption{\label{fig:Efield} Volume electric field amplitude distribution for TM$_{010}$ mode normalized to the stored energy of 1~J, which is used to calculate surface electric field participation.}
\end{figure}

The main challenge of measuring the ultra-high $Q > 10^{10}$ factor bulk SRF resonators at very low rf fields is the limited applicability of the standard continuous wave (CW) techniques~\cite{Melnychuk_RSI_2014} used for measuring at higher rf fields. In particular, power measurements are typically not bandpass filtered and are therefore limited by various sources of the rf noise present in the broad frequency range, and vector network analyzers do not have sufficient frequency stability to measure quality factors beyond $Q\sim 10^8$. We instead use single decay measurements~\cite{Reschke_SRF_93} of the transmitted power $P(t)$ with the additional narrow 10-10000~Hz bandpass filtering around the resonance to obtain $Q(E)$. Initially, the phase-locked loop keeps the cavity at resonance while the transmitted power is measured and a portion of it is directed through spectrum analyzer. Zero span measurements at the resonance frequency with a resolution bandwidth (RBW) of 10-10000~Hz are then performed by turning the RF field source off and capturing the time decay of $P(t)$. A feature of this technique is that it is insensitive to cavity frequency drift so long as the drift is less than the resolution bandwidth. At each moment in time $t_0$, the decay is described by the exponential:
\begin{equation}
P(t)=P(t_0)e^{-(t-t_0)/\tau}
\end{equation}
where $\tau$ is the instantaneous decay time constant, providing the direct measurement of the loaded quality factor ${Q_L(t)=\omega\tau}$. Using the input ($Q_1$) and pickup probe ($Q_2$) external quality factors obtained from CW calibration the unloaded quality factor $Q$ can then be calculated:
\begin{equation}
\frac{1}{Q}=\frac{1}{Q_L} - \frac{1}{Q_1} - \frac{1}{Q_2}
\end{equation}
Next, average surface resistance $R_\mathrm{s}=G/Q$ where $G=270$~$\Omega$ is the geometry factor obtained from electromagnetic field simulations can be obtained as a function of ${E_\mathrm{acc} \propto \sqrt{P(t)}}$. This methodology allows extending the lower boundary of rf fields at which $Q$ can be measured down to below $10^{-5}$~MV/m=10~V/m, or about $10^{12}$ photons on average. To obtain the intra-cavity photon number $n=U/h \nu$, we use simulations such as those shown in Fig.~\ref{fig:Efield} to calculate the stored energy, $U$, at a given accelerating field. The $Q$ measurement error is estimated to be lower than 10$\%$~\cite{Melnychuk_RSI_2014}.

Typical $P(t)$ data recorded using Rohde $\&$ Schwarz FSL-3N spectrum analyzer is shown in Fig.~\ref{fig:Pt} for the same cavity before (red curve) and after (black curve) the growth of $\sim$100~nm of additional surface oxide by anodizing, illustrating how instantaneous $\tau$ can be extracted, and how the differences in the dependence of $\tau (t)$ and therefore $Q (E)$ can be clearly observed. In this example, input and transmitted power couplings are similar for both curves, thus a faster decay with the much stronger time (rf field) dependence after anodizing (black curve) indicates additional strongly field dependent losses in the low field range. The curves have different noise floors due to the different resolution bandwidths and attenuations. 

\begin{table*}
\caption{\label{summary}Summary of results for investigated 1.3 GHz elliptical shape cavities.}
\begin{ruledtabular}
\begin{tabular}{*7c}
Cavity & Treatment & \multicolumn{2}{c}{$R_\mathrm{s}$(nOhm)} & $\Delta R_\mathrm{s}$(nOhm) & \multicolumn{2}{c}{TLS fit} \\
 &  & $5$~MV/m & $<0.001$~MV/m & & $E_\mathrm{c}$(MV/m) & $\beta$ \\ 
\hline
AES012 & Bulk EP & 2.7 & 9.0 & 6.3 & 0.19 & 0.38\\
AES012 & + 100 nm oxide by anodizing & 5.0 & 17.0 & 12.0 & 0.02 & 0.25\\
AES012 & + EP 5~$\mu$m & 3.0 & 7.0 & 4.0 & 0.19 & 0.38 \\
AES014 & Bulk EP + 120$^\circ$C 48 hrs & 2.6 & 8.6 & 6.0 & 0.14 & 0.41\\
AES015 & N infusion 800/120$^\circ$C 48 hrs & 2.0 & 5.2 & 3.2 & 0.21 & 0.33\\
AES015 & N infusion 800/160$^\circ$C 48 hrs & 1.8 & 4.4 & 2.6 & 0.18 & 0.29\\
RDTTD004\footnotemark[1] & N doping + condensed 10$^{-4}$~Torr of N$_2$ & 1.5 & 6.6 & 5.1 & 0.09 & 0.28 \\
AES011 & 800$^\circ$C 2 hrs +120$^\circ$C 48 hrs & 1.4 & 5.5 & 4.1 & 0.17 & 0.35\\
AES011 & N infusion 800/160$^\circ$C 96 hrs & 2.3 & 5.2 & 2.9 & 0.11 & 0.26 \\
AES016\footnotemark[1] & 800$^\circ$C 2 hrs +120$^\circ$C 48 hrs & 1.7 & 5.6 & 3.9 & 0.10 & 0.28  \\
PAV008\footnotemark[2] & 800$^\circ$C 3 hrs +120$^\circ$C 48 hrs & 9.8 & 17.0 & 7.2 & 0.12 & 0.37\\
PAV010 & N infusion 800/120$^\circ$C 48 hrs & 2.1 & 6.7 & 4.6 & 0.26 & 0.35 \\
PAV010 & N infusion 800/200$^\circ$C 48 hrs & 6.6 & 10.8 & 4.2 & 0.20 & 0.42 \\
\end{tabular}
\end{ruledtabular}
\footnotetext[1]{Large grain cavity, grain size of $\gtrsim 5$~cm}
\footnotetext[2]{This cavity had a higher than typical residual resistance at all fields, likely due to a manufacturing defect/inclusion}
\end{table*}%

Using both CW and single shot methods to extend the accessible field range we have measured $Q(E)$ dependencies for various 1.3 GHz fine grain ($\sim$50~$\mu$m) and large ($\gtrsim$5~cm) grain elliptical niobium cavities prepared by different surface treatments, including electropolishing (EP), EP+120$^\circ$C baking for 48 hours, nitrogen doping~\cite{Grassellino_SUST_2013}, and nitrogen infusion~\cite{Grassellino_SUST_2017}. Fast cooldowns with minimal ambient field to avoid flux trapping~\cite{Romanenko_APL_2014} were used in all cases. Measurements were performed around $T=1.5-1.6$~K where the contribution from thermally excited quasiparticles (typically referred to as `BCS') is small ($\lesssim$~1~nOhm for 1.3~GHz), and we therefore refer to the measured value as `residual'. In most cases, $T=2$~K measurements were also performed. 

\begin{figure}[htb]
 \includegraphics[width=\linewidth]{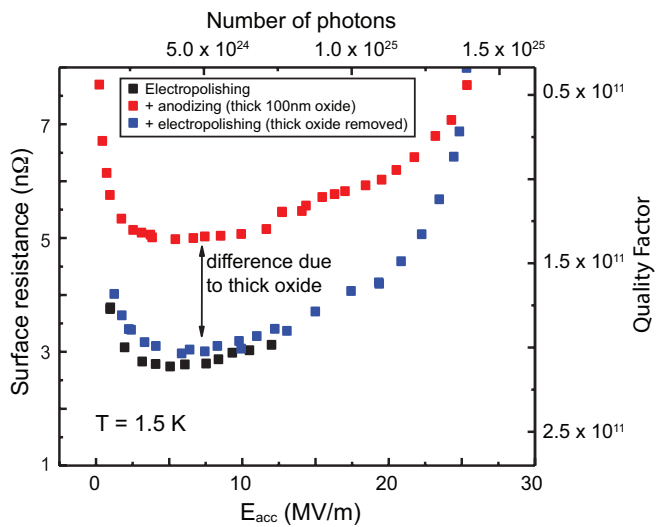}
\caption{\label{fig:Rres_Eacc_higher}Surface resistance as a function of the rf field amplitude measured at T=1.5-1.6~K. Removing the thick oxide by light electropolishing brings the residual resistance back down.}
\end{figure}

\begin{figure}[htb]
 \includegraphics[width=\linewidth]{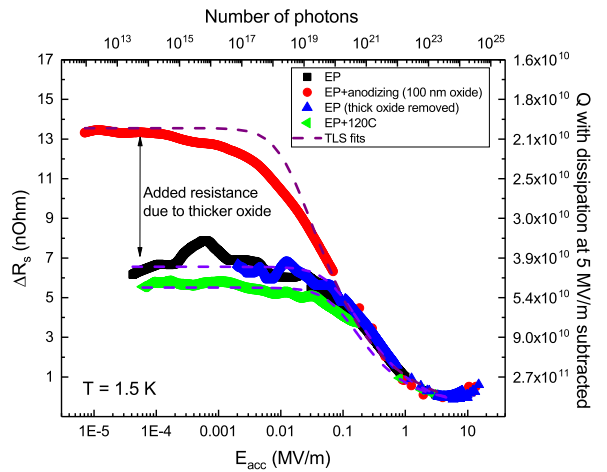}
\caption{\label{fig:Rres_Eacc}Additional surface resistance as compared with that found at 5~MV/m as a function of the rf rield amplitude measured at T=1.5-1.6~K. Dashed lines are fits to TLS model (Eq.~\ref{eq:TLS}).}
\end{figure}

The main finding of our work is shown in Fig.~\ref{fig:Saturation}: the low field $Q$ slope continues down to $E_\mathrm{acc}\sim$0.1~MV/m, and does not degrade further even at fields 1000x smaller. It is striking that this key finding was just slightly below the lowest fields $\sim$0.2~MV/m explored in previous studies of the low field $Q$-slope~\cite{Ciovati_JAP_2004}. All of the other cavities out of our large set prepared with different surface treatments (see Table~\ref{summary}) exhibited very similar $Q$ saturation behavior at low fields. The drop in $Q$ at higher fields of $\gtrsim$~3~MV/m is due to so-called medium and high field $Q$ slopes ~\cite{Romanenko_Rs_B_APL_2013, Padamsee_Ann_Rev_Nucl_2014}; non-equilibrium quasiparticle energy distribution driven by the rf field is another possible contributor~\cite{deVisser_PRL_2014}.

Interestingly, in 2D superconducting resonators, an increased low field dissipation has been known and studied for some time. Martinis~\emph{et. al.} proposed two-level systems (TLS) as the main cause of the increased low rf field losses in planar superconducting resonators~\cite{Martinis_PRL_2005}. Characteristic features of TLS have been well confirmed by subsequent experimental investigations~\cite{Gao_Thesis_2008}, and even individual TLS and interaction between them as a potential cause of the $1/f$ noise has been studied in detail lately~\cite{Burnett_NatureComm_2014, Lisenfeld_NatureComm_2015}. Two established signatures of the TLS-caused dissipation, which are directly relevant to our work are: 1) saturation of losses below a certain threshold; 2) dependence on the amount of the amorphous material exposed to the electric fields. The microscopic nature of TLS is believed to be due to individual atoms tunneling between two local energy minima within the amorphous part of the resonator, e.g. dielectric oxide between the electrodes in Josephson junctions, or a native oxide layer on the surface of superconductor. Some concrete microstructural candidates for TLS have also been described~\cite{Gordon_SciRep_2014} and identified experimentally~\cite{deGraaf_PRL_2017, Quintana_PRL_2017}. 

In the case of SRF cavities, amorphous niobium pentoxide layer of 3-5~nm is present on the inner cavity surface after all of the modern surface preparation techniques~\cite{Antoine_EUCARD_2012}. To probe if oxide is the origin of the increased low field dissipation, we selected one of the electropolished cavities with measured $Q(E)$ and grew a much thicker oxide of $\sim$100~nm by anodizing its inner surface using DC voltage of 48~V in the ammonia solution. This was followed by full $Q(E)$ measurements. We then removed thick oxide by electropolishing, and allowed the standard regrowth of the natural thin oxide layer. Residual (lower $T$) surface resistance for each of the three cases is shown in Fig.~\ref{fig:Rres_Eacc_higher}. Over the field range of 5-20~MV/m we observe an increase in the residual resistance of $\sim$2~nOhm.  Earlier studies~\cite{Palmer_SRF_1987} identified oxide as a contributor to the residual resistance at 5~MV/m and our results now show that the contribution remains about the same at higher fields.

Strikingly, the low rf field measurements reveal a much larger surface resistance increase. In Fig.~\ref{fig:Rres_Eacc} the residual surface resistance difference from its value at 5~MV/m is shown, and as much as 12~nOhm are added at $E_\mathrm{acc} \lesssim 0.01$~MV/m after anodizing. The comparison between different treatments is also shown in Table~\ref{summary}. Importantly, the surface resistance increase is fully reversed at all fields after the oxide is removed by EP. This experiment thus localizes the additional low field losses to niobium oxide, which is the second key finding of our work. 

We have also performed an additional experiment to probe if condensed gases on the surface of resonator may also affect the low field losses - the cooldown in the presence of $10^{-4}$~Torr of nitrogen inside the cavity - but found no observable change. 

Summary of all the measured average residual surface resistances at 5~MV/m and in saturation below 0.001 MV/m and a relative increase are shown in Table~\ref{summary}. We note that previously discussed models for LFQS~\cite{Palmieri_QSlopes_SRF2005, Halbritter_SRF_80} considered particular features \emph{within the magnetic penetration depth} rather than the surface oxide. From Table~\ref{summary} it follows that the structure of the penetration depth of niobium can be substantially modified (the treatments generate a variety of MFPs and defects), yet low field behavior is not significantly altered unless the dielectric surface oxide thickness is changed.

Saturation behavior and the role of oxide suggest that TLS may be a likely origin of the LFQS. According to the prevalent theory~\cite{Schickfus_PhysLettA_1977}, TLS-induced losses emerge from the dipole moments of `loose' atoms coupling to the electric field at the surface of resonators, and can be detected as an increased dielectric loss tangent $\delta_\mathrm{TLS}$. 

For resonators with TLS the $Q$ dependence on $E_\mathrm{acc}$ at low fields is considered~\cite{Martinis_PRL_2005, Wang_PRL_2009, Muller_arXiv_2017} to be of the form
\begin{equation}
\frac{1}{Q} = \frac{F \delta_\mathrm{TLS}(T)}{\left(1+\left(\frac{E_\mathrm{acc}}{E_\mathrm{c}(T)}\right)^2\right)^\beta} + \frac{1}{Q_\mathrm{qp}}
\label{eq:TLS}
\end{equation}
where $E_c$ is a characteristic electrical field for saturation, $Q_\mathrm{qp}$ is the non-TLS contribution of quasiparticles, $\beta$ is a fit parameter, and $F$ is the filling factor~\cite{Gao_APL_2008, Wang_PRL_2009, Wenner_APL_2011, Reagor_PRB_2016}, defined as
\begin{equation}
F = \frac{\int_{V_\mathrm{dielectric}}\epsilon_\mathrm{dielectric} |\vec{E (\vec{r})}|^2 d^3\vec{r}}{\int_{V_\mathrm{vacuum}}\epsilon_\mathrm{vacuum} |\vec{E (\vec{r})}|^2 d^3\vec{r}}.
\end{equation}
At $T \geq1.5$~K the temperature dependence of $\delta_\mathrm{TLS}(T)$ is likely residing in the plateau region~\cite{Enss_LTP}, thus we do not separate a $\tanh(h\nu/2kT)$ factor in Eq.~\ref{eq:TLS}, which is usually done for $T < 1$~K studies.

For TM$_{010}$ mode the distribution of the electric field over the cavity surface is not uniform, as obtained by COMSOL and CST Microwave Studio simulations shown in Fig.~\ref{fig:Efield}. For a 5~nm thick Nb$_2$O$_5$ layer with $\epsilon \approx 33$ we obtain $F \approx 3 \times 10^{-9}$, whereas for 100~nm after anodizing $F \approx 6 \times 10^{-8}$. The weighted contribution of TLS can be calculated as in~\cite{Wang_PRL_2009} and then used for fitting the observed $Q(E)$ dependencies to Eq.~\ref{eq:TLS}. Following this procedure, very good fits could be obtained, as shown in FIG.~\ref{fig:Saturation} and FIG.~\ref{fig:Rres_Eacc}. The best fit values of $\beta$ and $E_\mathrm{c}$ are listed in Table~\ref{summary}. The $\beta$ values range between 0.25 and 0.42 with no clear trend between different surface treatments. The lowest $E_\mathrm{c}$ value is obtained after anodizing, which may hint at a broader ensemble of TLS defects present in this case. Assuming that in saturation, $Q \sim 3 \times 10^{10}$ is dominated by TLS losses, we can obtain an estimate for $F \delta_\mathrm{TLS} \approx 1 / Q \approx 3 \times 10^{-11}$, which gives $\delta_\mathrm{TLS} \approx 10^{-2}$, which is close to what was measured in~\cite{Kaiser_SUST_2010}. We emphasize here that this value of $\delta_\mathrm{TLS}$ is likely corresponding to most of the TLS being thermally saturated.

It is interesting to note that the less pronounced LFQS in cavities at lower frequencies ($< 1$~GHz) is also consistent with our TLS hypothesis, as for $h \nu \ll k T$ the frequency dependence of $\delta_\mathrm{TLS}(\nu) \propto \nu$ makes the additional dissipation proportionally smaller. This has also been already shown experimentally in lumped-element resonators~\cite{Skacel_APL_2015, Muller_arXiv_2017}.

In summary, we observe the saturation of the low field $Q$ slope in bulk superconducting niobium cavities for particle accelerators at $E_\mathrm{acc} \lesssim 0.1$~MV/m, strongly increased low field dissipation for thicker surface oxide grown by anodization, and limited effect of treatments modifying the penetration depth but not the surface oxide. The high quality factor $Q > 10^{10}$ down to $\langle n \rangle \sim10^{12}$ photons provides promising outlook for using SRF cavities in the low-field applications. Our findings suggest that a likely cause of the LFQS may be - similarly to planar resonators - two-level systems in the natural niobium oxide, which may guide its mitigation. 

\acknowledgments
The authors would like to acknowledge technical support during measurements from O. Melnychuk and D. A. Sergatskov, and anodizing and electropolishing of some of the used cavities by A. Crawford. Fermilab is operated by Fermi Research Alliance, LLC under Contract No. DE-AC02-07CH11359 with the United States Department of Energy.

\end{document}